\documentclass{article}

\newcommand{\ag}{\hat{a}_{\gamma}}
\newcommand{\re}{\hat{a}^{\dagger}_{e}}

\newcommand{\rp}[1]{\hat{a}^{\dagger}_{\p_{#1}}}
\newcommand{\rpc}[2]{\hat{a}^{\dagger}_{\p_{#1#2}}}
\newcommand{\p}{\mathbf{p}}

\begin{document}

\title{Larc: a State Collapse Theory of Quantum Measurement}
\author{Michael Simpson}
\date{July 2010}   
\maketitle
\begin{abstract}
This proposes a new theory of Quantum measurement; a state reduction theory 
in which reduction is to the elements of the number operator basis of a system, triggered 
by the occurrence of annihilation or creation (or lowering or raising) operators in 
the time evolution of a system.  It is from these operator types that the acronym `LARC' is 
derived.  Reduction does not occur immediately after the trigger event; it occurs at 
some later time with probability $P_t$ per unit time, where $P_t$ is very small.  
Localisation of macroscopic objects occurs in the natural way: photons from an illumination 
field are reflected off a body and later absorbed by another body.  Each possible absorption 
of a photon by a molecule in the second body generates annihilation and raising operators, 
which in turn trigger a probability per unit time $P_t$ of a state reduction into the 
number operator basis for the photon field and the number operator basis of the electron 
orbitals of the molecule.  Since all photons in the illumination field have come from the 
location of the first body, wherever that is, a single reduction leads to a reduction of the 
position state of the first body relative to the second, with a total probability of $m\tau_L$, 
where $m$ is the number of photon absorption events.  Unusually for a reduction theory, 
the larc theory is naturally relativistic.  
\end{abstract}

\section{The Measurement Problem} 
Quantum Mechanical time evolution is
Hamiltonian time evolution. But quantum measurement, on the face of things,
appears to need a second law, unique to it. There are two reasons for this
apparent need. The first is that without a second law, measurements do not  
produce definite results. The second is that Hamiltonian evolution
is deterministic, but measurement outcomes are probabilistic; the
probabilistic element of quantum measurement is provided by the second law.
The most primitive form of second law is the `quantum jump'; a sudden,
probabilistic, change of state that occurs during a measurement.
Unfortunately, no-one has been able to convincingly specify the physical property of
measurement that invokes this special process. The measurement problem
consists in the apparent inability of pure Hamiltonian time evolution to
describe measurements correctly, combined with our inability to give any
special process a sound physical basis.  

	When using quantum mechanics for everyday calculation, physicists don't notice 
this problem, because they have learned to recognise physical arrangements that constitute 
measurement instruments, and at what point measurement can be said to have occurred, 
by applying the laws of quantum theory with what Bell referred to as `good taste'\cite{bell1}.  

	A solution to the measurement problem needs to either: 1) explain measurement
without recourse to special processes, or 2) specify the special processes as
physical laws, most importantly providing a law to explain when the process
occurs, since `measurement' is not physically well-defined.  This is a basis 
for a broad classification of solutions to the measurement problem: 
	type 1 solutions explain measurement without invoking any  
fundamental physical processes outside Hamiltonian time evolution;
	type 2 solutions specify new fundamental physical laws 
that allow measurement to be explained.  

	Everett's relative state interpretation is an example of a type 1 solution \cite{everett1}.  
Type 2 solutions can themselves be said to fall into two broad classes: those that 
retain the bulk of the quantum theoretical formalism, only adding the minimum needed 
to address the measurement problem itself; and those that depart more or less widely 
from the quantum formalism, producing effectively a new theory.  The Ghirardi-Rimini-Weber \cite{grw1} 
theory falls into the former class, while hidden-variable theories such as the Bohm\cite{bohm1} 
theory fall into the latter.  

	This present will propose a solution to the measurement problem of 
type 2, and of the former class of adding only enough to quantum theory to (with luck) 
explain the behaviour of measurement processes.  Maudlin\cite{maudlin1} has given a very 
succinct summary of the system of arguments that leads some to believe that the solution 
to the measurement problem must be of type 2.

	We are going to need at least one new law which must do two things.  It must 
specify when the state reduction process occurs, including what states of affairs 
trigger the process, and 
it must specify how the Hilbert space basis in which the reduction occurs is chosen.  Somewhere along 
the way, the theory must explain why this new process appears to occur only during 
measurements and not at other times.  Since there is no physical property possessed by 
measurement that distinguishes it from other processes, this is inherently tricky.  

\section{Outline of the Theory}  
Up to now, proposed type 2 solutions to the measurement 
problem have usually kept within nonrelativistic quantum mechanics.  They have often, even when 
adequate at the nonrelativistic level, met with difficulty in developing into 
a relativistic form\cite{grw1}\cite{bohm1}.  The event that will trigger state reduction in this theory, 
while it appears in nonrelativistic quantum mechanics, 
is more naturally associated with relativistic quantum mechanics and quantum field theories.  

	The ability of the number of particles, or of field quanta, to vary in quantum field theory is 
one of its fundamental properties.  Changes in the number of field quanta have operators 
associated with them, known as the annihilation ($\hat{a}$), and creation ($\hat{a}^\dagger$) operators.  
Processes in which a particle of a given type is annihilated and replaced by the creation of 
particles of other types are routine in quantum field theory.  Conservation laws demand that particle 
annihilation must be followed by particle creation.  (`Followed' here means in the temporal 
sense.  Formally, the creation operators fall to the left of the annihilation operators.)

	The concepts behind annihilation and creation operators also apply to the less 
esoteric case of a bound system in a certain energy state.  The classic case is the 
harmonic oscillator, for which there exist lowering ($\hat{a}$) and raising ($\hat{a}^\dagger$) operators that 
move the oscillator to the next lower or next higher energy state.  The harmonic oscillator 
is the classic example because it is one of the very few cases for which the mathematical 
form of the operators can be solved exactly.  However, in principle, corresponding operators exist for 
the electronic states of hydrogen atoms, or more complex molecules.  Lowering and raising 
operators mix with annihilation and creation operators, following the rule that a lowering or 
annihilation operator (or group of them) must be followed temporally by a raising or creation operator 
(or group of them) so that conservation laws are obeyed.  I'll refer to a well-formed cluster 
of lowering/annihilation - raising/creation operators as a larc.  

	One example of a larc would be pair creation: a photon annihilation operator followed 
by an electron creation operator and a positron creation operator.  Another example would be 
absorption of a photon by a molecule, described by a photon annihilation operator, followed by 
a raising operator for the electronic states of the molecule.  

	The larc theory postulates that the triggering event for a state reduction is the 
occurrence of a larc in the description of a process.  This does not mean that a state 
reduction happens at every larc event: a larc triggers a probability of a state reduction, 
specifically a probability per unit time.  If this probability per unit time is small 
enough, there will be virtually zero chance of a state reduction occurring during a 
process involving small numbers of larc events, which will be the case, for example, 
in high-energy particle experiments.  

	A state reduction consists of a lopping off of branches.  In most cases a larc event 
occurs as a branching: the larc event for absorption of a photon by a molecule will be a 
branch superposed with the alternative event of the photon passing through the molecule without 
being absorbed.  Each of these branches has an amplitude.  The presence of a larc in this 
process creates a probability per unit time that one of the branches will be lopped off, the 
probability for each branch to be the survivor is the square modulus of its amplitude.  

	In the larc theory, the basis in which the reduction occurs is the number basis 
for the particles or field quanta defined by the lowering/annihilation and raising/creation 
operators involved in each larc event.  

	In summary, the elements of the larc theory of quantum measurement are:
1.  State reduction occurs following a larc.  After a larc occurs there is a probability per unit 
time of a state reduction, which consists of a lopping off of the branches not chosen.  `Lopping 
off' means a zeroing of the amplitudes of those branches.  
2.  The basis in which the reduction occurs is the number basis defined by the operators in the 
larc.  

\section{Localisation of Macroscopic Bodies}  
So far, the larc theory doesn't appear to 
include anything that will solve the classic problem of quantum measurement: the localisation 
of macroscopic bodies.  Macroscopic body localisation is important for two reasons: first, it explains 
why the moon, 
or a cricket ball, is in a definite location (and is still there even when nobody looks).  Second, it 
ensures that the pointers of measuring devices must show definite outcomes.  Since 
pretty much any measuring device can be designed to show its output as the position of a pointer, 
or some equivalent, such as the position of macroscopic patches of colour on a display device, 
localisation of macroscopic objects is effectively a general solution to 
the measurement problem, at least to a first approximation.  

	That macroscopic objects are going to be localised by the larc theory may not be immediately 
obvious.  I've postulated a process that has nothing to do with position localisation, 
and occurs extremely rarely to boot.  In fact, the fundamental machinery required is already in 
place.  The completion of the argument to localisation requires a further property of quantum 
mechanics I will refer to as amplification by correlation (abc).  
As far as I know, this mechanism is due to Ghirardi, Rimini, and Weber\cite{grw1}.  

	In outline, abc 
acts whenever a composite state contains a large number of quanta that are in states correlated 
with one another.  The correlation 
means that if any one of the quanta gets reduced to a single definite state, all the other 
quanta in the composite state will be reduced to their corresponding definite states.  Since 
every quantum's state will be reduced if any quantum's state becomes reduced, the probability 
of reduction of the whole composite state becomes roughly the sum of the probabilities of 
each individual quantum being reduced.  In this way, the probability of reduction is amplified, 
potentially by many orders of magnitude, by the correlated state.  

	Body B begins in a superposition of positions $x_1$ and $x_2$.  Larc-abc is capable of 
localising a body whose position is unknown in three dimensions, but it is sufficient to show 
that it can localise B.  The space where B exists can be illuminated.  Initially, the light is 
off.  At $t=t_i$ the light will turn on.  

	The final element of the thought experiment is a pinhole camera, C.  This is 
positioned in the plane, facing towards the line we know B is confined to.  C consists of 
a box with a small hole in the side facing B (not small compared to the wavelength of the 
illumination, of course).  At the back of the camera is an old-fashioned photographic plate.  

\begin{picture} (100,200)(-10,-20)
\thicklines
\put(0,25)
{
\put(35,15){B}
\put(217,45){C}
%\put(200,60){\framebox(40,40)}
\put(200, 60.35) {
\put(0,0){\line(0,1){18}}
\put(0,22){\line(0,1){18}}
\put(0,0){\line(1,0){40}}
\put(0,40){\line(1,0){40}}
\put(40,0){\line(0,1){40}}
}
\put(28,118){$x_1$}
\put(240,70){\line(-4,1){200}}
\put(244,66.5){$\xi_1$}
\put(28,38){$x_2$}
\put(240,90){\line(-4,-1){200}}
\put(244,87){$\xi_2$}
}
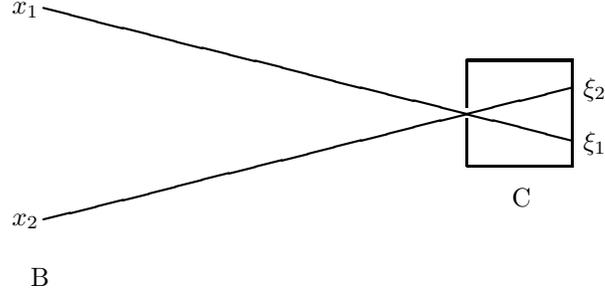
\put(45,0){Localisation of body B by pinhole camera C}
\end{picture}

	Let's start with the simplest possible case; B in a superposition of two possible 
locations, $x_1$ and $x_2$.  
\begin{equation}
\psi_{B} = a_{1} \delta (x-x_1) + a_{2} \delta (x-x_2)
\label{e1}
\end{equation}
With $|a_{1}|^{2} + |a_{2}|^{2} = 1$.  
At time $t_{i}$, the light turns on
\begin{equation}
a_{1} \delta (x-x_1) I(x,t_{i}) + a_{2} \delta (x-x_2) I(x,t_{i})
\label{e2}
\end{equation}
Where $I(x,t_{i})$ represents the electromagnetic field radiation expanding from point 
$x$, at time $t_{i}$.  

	The photographic plate is composed of individual molecules capable of absorbing visible photons 
and, as a consequence, undergoing a change which will later be visible to the 
unaided human eye, provided enough molecules absorb photons.  At time $t_f$ the plate has been 
illuminated by $n$ photons, where $n$ is presumed to be large enough 
to ensure that the result will be visible.  

	Again to simplify the mathematics, and without loss of generality, 
we assume that the camera is positioned equidistant from $x_1$ and $x_2$, so that the number of photons 
illuminating the plate is the same for either possible position of B.  If the body B is at $x_1$, then a patch 
of the photographic plate will be illuminated, call this patch $\xi_1$.  If the body is at $x_2$, a different 
patch will be illuminated, call it $\xi_2$.  The illumination landing on $\xi_i$ at time $t_f$ will be represented 
by $I(\xi_i,t_f,n)$; $n$ is the number of photons that will fall on the patch $\xi_i$.  

	As $I$ passes through each molecule it generates a 
factor representing possible absorption of a photon by a molecule at the photographic plate.  A very simple 
representation of this factor is 
\begin{eqnarray}
\lefteqn{
a_{1} b_{1} \delta (x-x_1) \ag I(\xi_{1},t_f,n) \re \psi(\xi_{1},i) \psi(\xi_{2},i)
} \nonumber \\ & \mbox{}
+ a_{1} b_{2} \delta (x-x_1) I(\xi_{1},t_f,n) \psi(\xi_{1},i) \psi(\xi_{2},i)
\nonumber \\ & \mbox{}
+ a_{2} b_{1} \delta (x-x_2) \ag I(\xi_{2},t_f,n) \psi(\xi_{1},i) \re \psi(\xi_{2},i)
\nonumber \\ & \mbox{}
+ a_{2} b_{2} \delta (x-x_2) I(\xi_{2},t_f,n)) \psi(\xi_{1},i) \psi(\xi_{2},i) 
\label{e3}
\end{eqnarray}
%\begin{eqnarray}
%\lefteqn{
%b_{2} (a_{1}I(\xi_{1},t_f,n) + a_{2}I(\xi_{2},t_f,n)) \psi(\xi_{1},i) \psi(\xi_{2},i) 
%} \nonumber \\ & &
%+ a_{1} b_{1} \ag [I(\xi_{1},t_f,n-1)] \aae \psi(\xi_{1},i) \re \psi(\xi_{1},f) \psi(\xi_{2},i)
%\nonumber \\ & &
%+ a_{2} b_{1} \ag [I(\xi_{2},t_f,n-1)] \psi(\xi_{1},i) \aae \psi(\xi_{2},i) \re \psi(\xi_{2},f)
%\label{e3}
%\end{eqnarray}
	Where $|b_{1}|^{2} + |b_{2}|^{2} = 1$, $b_{1}$ being the amplitude for a photon to be 
absorbed by a light-sensitive molecule, while $b_{2}$ is the amplitude for the photon to 
pass through without being absorbed (in a given time).  $\psi(\xi_{1},i)$ is the initial electron 
orbital state of a light-sensitive molecule in area $\xi_{1}$, the area on the photographic plate 
that will be illuminated if B is at $x_{1}$.  $\ag$ is a photon annihilation operator such that 
$\ag I(\xi_{1},t_f,n) = n^{1/2} I(\xi_{1},t_f,n-1)$.  $\re$ is 
a raising operator acting on the electron orbital of the molecule, raising it to the energy 
state that leads to a visible change, so that $\re \psi(\xi_{1},i) = \psi(\xi_{1},f)$.  

	When states of the form (\ref{e3}) exist the proposed state reduction mechanism becomes active.  
The mechanism has three elements.  

	(1) Trigger.  Once the state contains a larc there is a probability per unit time of a reduction.  
Call the time constant for this $P_t$.  
In this reduction branches other than the one selected will be lopped off, their amplitudes set to zero.  

	(2) Basis choice.  The basis in which this reduction 
occurs is the number basis consistent with the annihilation and raising operators that appear in 
the larc used to describe the absorption of the photon by the molecule.  In this case the 
relevant number operators are $\hat{n}_\gamma = \ag^\dagger \ag$, corresponding to the number of photons 
in the incoming electromagnetic field, $\hat{n}_{ei}$
the number of electrons in the initial electron orbital state, and $\hat{n}_{ef}$, the number of electrons in 
the final electron orbital state.  

	I will express these combinations of number operator eigenvalues in the form of a 
3-tuple $(n, i, f)$, where $n$ is the number of photons, $i$ the number of electrons in the initial state, and 
$f$ the number of electrons in the final state.  $i$ and $f$ can each only be zero or one, because the electron 
is a fermion, and $i + f = 1$, because the number of electrons is conserved.  The initial state is 
$(n, 1, 0)_1$, where the subscript indicates which of the two regions, $\xi_1$ or $\xi_2$, is in question.  
There are three possible final combinations: if the photon is absorbed the final state is 
$(n-1, 0, 1)_1$, which 
corresponds to a zeroing of amplitude $b_{2}$; if the photon is not absorbed the state is 
$(n, 1, 0)_1$, which corresponds to a zeroing of amplitude $b_{1}$.  In each of these cases it is inherent 
that $a_{2}$ is also zeroed.  

	There remains a third possible final 
combination: $(n-1, 1, 0)_1$.  This corresponds not to the photon being absorbed by the molecule, but to 
a removal of one photon from the electomagnetic field $I(\xi_{1},t_f,n)$ without its being absorbed by 
the molecule at $\xi_{1}$.  On the face of things, this outcome violates 
energy conservation, since the photon disappears but its energy does not get transferred to the electron 
orbital of the molecule.  However, recall that (\ref{e3}) is a superposition of two different 
positions for B, and consequently, two different illumination states.  If a photon disappears from 
one of the two superposed illumination states, then it still exists in the other.  This outcome 
corresponds with a zeroing of $a_{1}$.  

	This third way of reducing the state is necessary to the success of the whole enterprise.  
Imagine an alternative version of the pinhole camera, in which there is film in only the half of 
the camera containing $\xi_1$, the half containing $\xi_2$ being open, so that any photons that would 
have been absorbed at $\xi_2$ now pass through empty space without being absorbed at all.  If in this 
situation only the first two combinations were allowed, then with probability one, B would be at $x_1$.  
It would be possible to determine where B is simply by choosing whether to detect the illumination beam 
at $\xi_1$ or $\xi_2$.  Therefore, combination 3 is necessary if the larc-abc mechanism is to produce 
the correct Born rule probabilities.  

	(3) State Reduction.  The final result of the state reduction process is one of three possible states.  The first 
in which the photon is definitely absorbed, is that corresponding to $(n-1, 0, 1)_1$

\begin{equation}
\delta (x-x_1) \ag I(\xi_{1},t_f,n) \re \psi(\xi_{1},i) \psi(\xi_{2},i)
\label{e3.2}
\end{equation}
which occurs with probability $|a_{1} b_{1}|^2$.  Of course, this state is equal to 
\begin{equation}
\delta (x-x_1) n^{1/2} I(\xi_{1},t_f,n-1) \psi(\xi_{1},f) \psi(\xi_{2},i)
\label{e3.3}
\end{equation}

The second, in which the photon is definitely not absorbed, corresponding to $(n, 1, 0)_1$

\begin{equation}
\delta (x-x_1) I(\xi_{1},t_f,n) \psi(\xi_{1},i) \psi(\xi_{2},i)
\label{e3.4}
\end{equation}
which occurs with probability $|a_{1} b_{2}|^2$.

	The third case is a little harder to describe neatly.  One photon is removed from the illumination 
field in the $\xi_{1}$ region, without being absorbed.  This situation corresponds to $(n-1, 1, 0)_1$

\begin{eqnarray}
\lefteqn{
b_{1} \delta (x-x_2) \ag I(\xi_{2},t_f,n) \psi(\xi_{1},i) \re \psi(\xi_{2},i)
} \nonumber \\ & \mbox{}
+ b_{2} \delta (x-x_2) I(\xi_{2},t_f,n)) \psi(\xi_{1},i) \psi(\xi_{2},i) 
\label{e3.6}
\end{eqnarray}
which occurs with probability $|a_{2}|^2$.  

\section{Amplification By Correlation}
	As exposure goes on, we get large numbers of factors like (\ref{e3}) representing possible 
absorption 
of photons by the molecules in regions $\xi_{1}$ and $\xi_{2}$.  Assuming that the photographic 
plate's coating is opaque, all the photons will eventually be absorbed.  After the entire 
illumination front has passed the camera, the state at the photographic plate 
will look something like
\begin{eqnarray} 
\lefteqn{ \textstyle{
a_{1} \delta (x-x_1) \prod_{j=1}^{m_{1}}\left[b_{2}\psi_{j}(\xi_{1},i) + b_{1}\re\psi_{j}(\xi_{1},i)\ag
\right]I(\xi_{1},t_f,n) 
}} \nonumber \\ & \textstyle{ \mbox{} 
\times \prod_{j=1}^{m_{2}}\psi_{j}(\xi_{2},i)
}\nonumber \\ &  \textstyle{  \mbox{} 
+ a_{2} \delta (x-x_2) \prod_{j=1}^{m_{2}}\left[b_{2}\psi_{j}(\xi_{2},i)+b_{1}\re\psi_{j}(\xi_{2},i)\ag
\right]I(\xi_{2},t_f,n) 
} \nonumber \\ &  \textstyle{ \mbox{} 
\times \prod_{j=1}^{m_{1}}\psi_{j}(\xi_{1},i)
}\label{e4}
\end{eqnarray}
%\begin{eqnarray}
%\lefteqn{
%&
%a_{1} \delta (x-x_1) \prod_{j=1}^{m_{1}}\left[ b_{2}\psi_{j}(\xi_{1},i)I(\xi_{1},t_f,n) + b_{1} \aae\psi_{j}(\xi_{1},i) %\re\psi_{j}(\xi_{1},f) \ag
%I(\xi_{1},t_f,n-m_1) \right]
%\nonumber \\ & 
%\prod_{j=1}^{m_{2}}\psi_{j}(\xi_{2},i)
% \nonumber \\ &
%+ a_{2} \delta (x-x_2) \prod_{j=1}^{m_{2}}\left[ b_{2}\psi_{j}(\xi_{2},i)I(\xi_{2},t_f,n) + b_{1} \aae\psi_{j}(\xi_{2},i) %\re\psi_{j}(\xi_{2},f) \ag
%I(\xi_{2},t_f,n-m_2) \right]
%\nonumber \\ & 
%\prod_{j=1}^{m_{1}}\psi_{j}(\xi_{1},i)
%\label{e4}
%\end{eqnarray}
Where $m_1$ and $m_2$ represent the number of absorbing molecules in regions 1 and 2 respectively.  

	As before there are three elements to the state reduction process, and three possible end 
states.  

	(1) Trigger.  The state described by equation (\ref{e4}) contains a large number of larc events.  Each one creates 
a probability per unit time of a state reduction.  The probability per unit time is $P_t$ for each 
larc event.  The total number of larc events in (\ref{e4}) is $m_1 + m_2$, which means that, roughly, the 
probability per unit time for a reduction is $P_t(m_1 + m_2)$.  

	(2) Basis Choice.  The basis choice for any indidvidual larc event is the same as that for equation (\ref{e3}).  
Let us say, without loss of generality, that the larc event associated with molecule $q$ in patch $\xi_1$ is the 
first to trigger a state reduction.  The basis will then be that associated with the photon number operator, 
the number operator for the initial state of molecule $1;q$, and the number operator for the final state of 
molecule $1;q$.  The possible outcomes, characterised as above, are $(n-1,0,1)_{1;q}$, $(n,1,0)_{1;q}$, and 
$(n-1,1,0)_{1;q}$.  

	(3) State Reduction.  Again, one of three possible states results from the process.  The first, 
$(n-1,0,1)_{1;q}$, corresponds to the photon being definitely absorbed at $1;q$.  

\begin{eqnarray}
%\lefteqn{
\delta(x-x_1) \re\psi_{q}(\xi_{1},i)\ag \prod_{\stackrel{\scriptstyle j=1}{j\ne q}}^{m_{1}}\left[b_{2}\psi_{j}(\xi_{1},i) 
+ b_{1}\re\psi_{j}(\xi_{1},i)\ag \right]I(\xi_{1},t_f,n) 
%}  
\nonumber  \\ \mbox{}
\times \prod_{j=1}^{m_{2}}\psi_{j}(\xi_{2},i) 
\label{e4.2}
\end{eqnarray}
with probability $|a_1 b_1|^2$.

	The core of the larc-abc mechanism is now clear.  
The crucial property of the state described by (\ref{e4}) is this: the photons are all 
correllated.  If any one photon is 
absorbed in region $\xi_{1}$, then every photon must be absorbed in region $\xi_{1}$ and none in region 
$\xi_{2}$.  This is clear in (\ref{e4.2}), which is the result of a single photon definitely absorbed in 
region $\xi_{1}$.  All terms in which a photon could be absorbed in region $\xi_{2}$ have vanished.  
Since the region in which a photon is absorbed is correlated with the position of B, 
any photon absorbed in $\xi_{1}$ means the position state of B is $\delta (x-x_1)$, which corresponds to 
a zeroing of $a_2$.  Again, this is clear in (\ref{e4.2}): all the terms with $\delta (x-x_2)$ are gone.  

	The second possible end state after a state reduction is $(n,1,0)_{1;q}$, corresponding to a photon 
being definitely not absorbed at $1;q$.  

\begin{eqnarray}
%\lefteqn{
\delta (x-x_1) \psi_{q}(\xi_{1},i) \prod_{\stackrel{\scriptstyle j=1}{j\ne q}}^{m_{1}}\left[b_{2}\psi_{j}(\xi_{1},i) 
+ b_{1}\re\psi_{j}(\xi_{1},i)\ag \right]I(\xi_{1},t_f,n) 
%} 
\nonumber  \\*  \mbox{}
\times \prod_{j=1}^{m_{2}}\psi_{j}(\xi_{2},i)
\label{e4.4}
\end{eqnarray}
with probability $|a_1 b_2|^2$.  

	The third end state is $(n-1,1,0)_{1;q}$, corresponding to the photon being removed from the illumination field at 
$\xi_1$ without being absorbed at $1;q$.  
\begin{eqnarray}
%\lefteqn{
%a_{1} \delta (x-x_1) \prod_{j=1}^{m_{1}}\left[b_{2}\psi_{j}(\xi_{1},i) + b_{1}\re\psi_{j}(\xi_{1},i)\ag
%\right]I(\xi_{1},t_f,n) 
%} \nonumber \\ & 
%\prod_{j=1}^{m_{2}}\psi_{j}(\xi_{2},i)
%\nonumber \\ &    \mbox{}
\delta (x-x_2) \prod_{j=1}^{m_{2}}\left[b_{2}\psi_{j}(\xi_{2},i)+b_{1}\re\psi_{j}(\xi_{2},i)\ag
\right]I(\xi_{2},t_f,n) 
 \nonumber \\*   \mbox{}
\times \prod_{j=1}^{m_{1}}\psi_{j}(\xi_{1},i)
\label{e4.6}
\end{eqnarray}
with probability 
$|a_2|^2$. 

	Effectively, B has been localised to $\delta (x-x_2)$ before any photon 
has been definitely absorbed, or definitely not absorbed, at $\xi_2$.  
The form of equation (\ref{e4.6}) is determined by the correlation of photons in the illumination field 
$a_{1} I(\xi_{1},t_f,n) + a_{2} I(\xi_{2},t_f,n)$.  The photons are correlated such that they are either 
all at $\xi_1$, or all at $\xi_2$.  

	So, equations (\ref{e4.2}), (\ref{e4.4}), and (\ref{e4.6}), are the final end results of the measurement 
process.  In each of them, body B has become definitely localised.  
As I have said, if even a single photon undergoes a state reduction, it carries all the photons with it, 
as well as the body B with whose position they are all in turn correlated.  This means that even if the 
probability per unit time of a reduction is very small for each larc event, the enormous number of 
larc events involved implies that the probability of B becoming localised may be virtually one.  
This is the effect I have dubbed amplification by 
correlation (abc): a very small probability of a state reduction ($P_t$) for each photon is amplified by 
correlation of large enough numbers of photons ($m_1 + m_2$), into a very large probability of reduction of the 
position of body B ($P_t(m_1 + m_2)$).  Hence (\ref{e4}) implies that B becomes 
localised with probability virtually one.  

	It is worth being clear about what is and is not correlated in (\ref{e4}).  If any photon 
is definitely absorbed at $\xi_1$, then no photon can be absorbed at $\xi_2$, all must ultimately 
be absorbed at $\xi_1$.  However, while the dropping of one molecule into  a definite state forces the 
position of B into a definite state, it does not force any of the other photons to decide whether 
they have been definitely absorbed.  As a result, it is possible that when a person looks at the 
plate, there will not be any definite fact of the matter of exactly which molecules have been changed 
and which have not.  Some molecules may still be in a superposition of having absorbed or not absorbed 
a photon.  But this is not as serious as it might sound.  The differing states that are superposed are 
not distinguishable to the human observer because they all lie within one region   Furthermore, it is 
likely that photon absorption and amplification processes that occur within the visual system and brain 
will raise the probability of these superposed states becoming definite.  

	It is clear that the above argument can be generalised to a body whose position is unknown in 
one, two or three dimensions, using two or three pinhole cameras.

\section{Human-independent localisation}
	The argument so far has relied on the pinhole camera, which is clearly a human artifact.  
Can we expect that the larc-abc mechanism described above will lead to localisation by other 
measuring devices, or even without 
any human involvement?  The answer is yes, for more than one reason.  

	First, the involvement of humans, as observers, is 
irrelevant to the localisation process.  The only involvement of humans as observers (as opposed to 
artificers) in the above argument is the specification that the state reached by a molecule when it 
absorbs a photon, $\re \psi(\xi_{1},i)=\psi(\xi_{1},f)$, is visibly different, to the human eye, when compared with the 
initial state, $\psi(\xi_{1},i)$.  But this assumption plays no role in the argument leading to 
localisation.  Localisation would occur in exactly the same way if the two molecular states were 
not visibly different to the human eye, and hence does not depend on the camera's use as a 
measuring instrument, but only on its basic physical characteristics.  

	Second, localisation will occur even if the `pinhole camera' is 
just a natural cavity in the surface of a second body.  After all, 
real macroscopic bodies are not smooth surfaced; real bodies have surfaces 
which have many ridges and valleys, hills and cavities.  At the bottoms of these valleys and cavities 
will be places where illumination from another body can reach only if it is in the right relative 
position, so the above analysis based on the `pinhole camera' will still apply.  
%
%\begin{picture} (100,200)(-10,-20)
%\thicklines
%\put(35,10){B}
%\put(217,45){C}
%\put(200,60){\framebox(40,40)}
%\put(20,25){\begin{picture}(60,60)
%  \qbezier(125,57.5)(125,60)(135,60)
%  \qbezier(125,57.5)(125,55)(135,55)
%  \qbezier(145,52.5)(145,55)(135,55)
%  \qbezier(145,52.5)(145,50)(135,50)
%
%  \qbezier(125,46)(125,50)(135,50)
%  \qbezier(125,46)(125,47)(135,47)
%  \qbezier(145,48)(144,48)(135,47)
%  \qbezier(145,48)(146,40)(135,40)
%\end{picture}}
%
%\put(50,55){\begin{picture}(60,60)
%  \qbezier(125,70)(125,75)(135,75)
%  \qbezier(125,70)(125,65)(135,65)
%  \qbezier(145,60)(145,65)(135,65)
%  \qbezier(145,60)(145,55)(135,55)
%
%  \qbezier(125,50)(125,55)(135,55)
%  \qbezier(125,50)(125,45)(135,45)
%  \qbezier(145,42.5)(145,45)(135,45)
%  \qbezier(145,42.5)(145,40)(135,40)
%\end{picture}}
%
%\put(28,118){$x_1$}
%\put(240,70){\line(-4,1){200}}
%\put(244,66.5){$\xi_1$}
%\put(28,38){$x_2$}
%\put(240,90){\line(-4,-1){200}}
%\put(244,87){$\xi_2$}
%\end{picture}

	Third, and more generally, when a photon is absorbed by a molecule or crystal lattice it 
transfers information 
about the direction of its source in the form of a momentum kick.  Imagine a single molecule, M, part of 
a macroscopic body positioned somewhere off the line $x$, which is struck by the illumination field 
coming from body B.  
The illumination is a superposition of illumination fields coming from two different directions.  
Direction one corresponds to a photon coming from position $x_1$, which will give the molecule a kick $\p_1$.  
Direction two corresponds to a photon coming from position $x_2$, which will give the molecule a kick $\p_2$.  
Once the illumination field has passed through the molecule the state will be 
\begin{eqnarray}
\lefteqn{
a_1 b_1 \delta (x-x_1) \ag I(\p_{1},t_f,n) \re \psi(i) \rp{1} \phi(\p_i) 
} \nonumber \\ & & \mbox{}
+ a_1 b_2 \delta (x-x_1) I(\p_{1},t_f,n) \psi(i) \phi(\p_i)
\nonumber \\ & & \mbox{}
+ a_2 b_1 \delta (x-x_2) \ag I(\p_{2},t_f,n) \re \psi(i) \rp{2} \phi(\p_i)
\nonumber \\ & & \mbox{}
       + a_2 b_2 \delta (x-x_2) I(\p_{2},t_f,n) \psi(i) \phi(\p_i)
\label{e5}
\end{eqnarray}
where $\p_1$ ($\p_2$) represents the momentum of a photon travelling from $x_1$ ($x_2$) towards M.  It is no longer the 
position of the molecule M that is correlated with the position of B; the position of B is now correlated 
with the direction of the momentum striking a single molecule.  
$\phi(\p_i)$ is the centre of mass momentum state for the molecule M, with an initial momentum of $\p_i$.  
$\rp{1}$ is a raising operator affecting the centre of mass momentum state of the molecule and 
adding momentum $\p_1$ to it; hence $\rp{1} \phi(\p_i) = \phi(\p_i + \p_1)$.  

\begin{picture} (100,200)(-10,-20)
\thicklines
\put(0,20)
{
\put(35,10){B}
\put(196,45){M}
%\put(200,60){\framebox(40,40)}
\put(200,80){\circle*{5}}
\put(28,118){$x_1$}
\put(200,80){\line(-4,1){160}}
\put(200,80){\vector(4,-1){50}}
\put(254,65.5){$\p_1$}
\put(28,38){$x_2$}
\put(200,80){\line(-4,-1){160}}
\put(200,80){\vector(4,1){50}}
\put(254,90){$\p_2$}
}
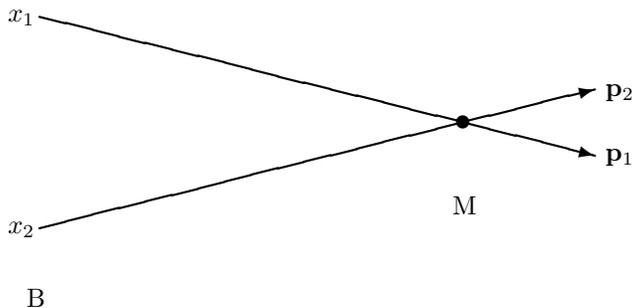
\put(10,0){Localisation of body B by momentum transferred to molecule M}
\end{picture}

	Decomposing the momentum kick into components 
$\p_1 = \p_{1x} + \p_{1y}$, we have 
\begin{eqnarray}
\lefteqn{
a_1 b_1 \delta (x-x_1) \ag I(\p_{1x},\p_{1y},t_f,n) \re \psi(i) \rpc{1}{x} \rpc{1}{y} \phi(\p_{ix} + \p_{iy}) 
} \nonumber \\ & & \mbox{}
+ a_1 b_2 \delta (x-x_1) I(\p_{1x},\p_{1y},t_f,n) \psi(i) \phi(\p_i)
\nonumber \\ & & \mbox{}
+ a_2 b_1 \delta (x-x_2) \ag I(\p_{2x},\p_{2y},t_f,n) \re \psi(i) \rpc{2}{x} \rpc{2}{y} \phi(\p_{ix} + \p_{iy})
\nonumber \\ & & \mbox{}
+ a_2 b_2 \delta (x-x_2) I(\p_{2x},\p_{2y},t_f,n) \psi(i) \phi(\p_i)
\label{e6}
\end{eqnarray}
The larcs in equation (\ref{e6}) create a probability per unit time $P_t$ of a state reduction which will leave only 
one term of equation (\ref{e6}) behind.  If we consider all the molecules exposed to the illumination field 
$I(n)$ we will end up with a state like (\ref{e4}), but one in which the direction of the momentum transferred 
to each molecule is correlated, rather than the position in which the transfer takes place, which means that 
the abc mechanism will apply to the momenta transferred by many photons to a macro body: the directions 
of these momenta will be correlated, and just one reduction to a definite direction will carry all the 
others with it through this correlation.  Localisation of body B occurs without any human involvement, 
and with no assumptions about the structure of the other body(s).

%	It is worth noting that the use of amplification by correlation may not be vital to the theory.  

\section{Some properties of the theory}  
	A significant property of the theory is that while it localises 
macroscopic objects, it can only produce localisation relative to other bodies.  The larc theory is 
in principle incapable of producing absolute position localisation.  I regard this as a strength of the theory.  
Most physical theories that are not explicitly relativistic are agnostic about the relativity of 
position: position variables can be interpreted as either relative or absolute without any effect on 
observable physics.  The larc-abc theory is a rare example of a theory that permits only of relative 
positions.  

	Type 2 theories of quantum measurement are usually formulated initially in nonrelativistic 
terms, and they often have problems extending to the relativistic case.  Both the Bohm\cite{bohm1} and 
GRW\cite{grw1} theories have difficulties of this kind.  
The larc theory, by contrast, analyses the situation the same way it would be done in quantum electrodynamics.  
Larc clusters 
are already present in quantum field theory, so they do not disrupt the relativistic invariance of the 
analysis.  The branches in the larc number operator basis for bound states effectively define the energy basis, which 
again is consistent with a relativistic theory.  If each branch is relativistic, then lopping off any 
one branch, or all but one, will leave behind a system that remains relativistic.  

	It is a vital property of the larc-abc process that it leads to position localisation in appropriate 
circumstances.  But this localisation can only occur relative 
to some other body.  Therefore, the theory cannot produce any problem of there being a preferred reference frame.  
Likewise, the temporal constant $P_t$ for the probability of a state reduction also belongs to a well-defined 
system: that to which the relevant larc cluster belongs, such as the molecule where the photon is absorbed, 
so again there is no need for a preferred reference frame.  Taken altogether, larc theory should be as 
relativistic as quantum field theory itself.  

	Similar arguments show that the theory conserves energy.  The emission of photons from B conserves 
energy, the absorption of photons, either in the pinhole camera, or in molecule M, also conserves energy.  
Each branch in the superposition conserves energy, so if the amplitudes of all but one branch are zeroed, 
whichever branch remains conserves energy.  

	Larc theory has a property which could be considered a further advantage over the GRW theory.  
GRW theory has two free parameters: the 
probability per unit time for a spontaneous localisation, and the parameter defining the width of the localised 
state.  In fact, the aspect of the theory that is free to be adjusted is much larger than two parameters, because 
the position distribution after spontaneous localisation is free; a wide range of arbitrary localised distributions can
be chosen.  

	Larc theory does not have this freedom: the distribution of the localised state is determined by 
the detail of the mechanism that establishes the correlation between relative position and the bound electron 
states within molecules.  In fact, the distribution of the localised state is determined in exactly the same 
way it would have been in standard 'good taste' quantum theory.  

	The time distribution 
for the spontaneous localisation in GRW is also free, but in GRW a constant probability per unit time is natural 
given that every particle may localise at any time.  The only event that could form an anchor for a distribution 
not constant in time would be each particle's own localisations, or possibly those of particles spatially 
correlated with the first.  In this exposition I have assumed that the Larc theory also 
produces a constant probability per unit time.  Although the justification for this is less natural, because 
Larc theory does have a trigger event relative to which the probability distribution could be defined.  Nevertheless, 
constant probability per unit time remains a natural expectation in the Larc theory.  

So Larc theory contains only one free parameter, where GRW has two, plus significant freedom in the position 
distribution.  

	I have not made any attempt to resolve one point.  Larc clusters occur also in virtual processes.  
There is therefore the question of whether virtual larc clusters should be considered to trigger state 
reductions just as real ones do.  If they do, it is also not clear whether the temporal constant should be 
the same in both cases.  I have no basis on which to offer an opinion on whether virtual larc clusters should be 
considered as 
triggering state reductions.  The only comment I have to make is that, if they do, this will certainly have an effect 
on the calculation of virtual processes, but with results that may be significant only in those of high complexity.

\section{Empirical tests}
	Since the core purpose of a measurement theory is to match the well-known behaviour of measurements, 
without changing the known behaviour of other types of quantum systems, 
empirical verification of a measurement theory is inherently difficult.  Nevertheless, 
as in any other case where one has changed the fundamental laws underlying a theory, one expects 
that there will be at least one situation where the predictions of the new theory diverge from those of the old.  
This is what we expect for all type 2 solutions to the measurement problem.  Measurement is not a 
physically well-defined concept, so it is next to impossible that any physical law could exactly match its 
behaviour.  At some point, there will be a situation where the new law operates detectably when there is 
no measurement occurring.  

	The problem of verification is made worse by the phenomenon usually called environment-induced 
decoherence.  This phenomenon mimics an aspect of the behaviour of measurements by causing the branches 
corresponding to the possible outcomes of measurements to lose their coherence.  A measurement 
necessarily requires the measured system to be put into correlation with the measuring apparatus.  
If one considers the measuring apparatus by itself, tracing out the measured system, its possible outcome 
states become decohered.  As the 
total state of the system and apparatus remains a superposition this does not solve the measurement 
problem, but it does mean that interference effects that would be destroyed by a successful measurement 
theory are destroyed in any case by decoherence, making straightforward tests impossible.  

	Nevertheless, the possibility of processes that generate large enough numbers of larcs to produce 
detectable loss of interference, without its being masked by environment-induced decoherence, remains.  

	If larc-triggered decoherence occurs during virtual processes, this may also produce detectable 
deviation from behaviours predicted by standard quantum field theory.

\section{Conclusion}  
	In 1990, Shimony\cite{shim1} provided a heuristic justifiction for a theory in which state 
reductions result from sudden jumps to energy eigenstates.  Technically, larc theory is not such a 
theory, since in it jumps are to eigenstates of number operators defined by larc operators, not energy 
operators.  Nevertheless, 
since the number operator states for electron orbitals correspond to bound state energy eigenstates, larc 
theory is essentially a theory of the type Shimony called for.  Shimony's argument 
therefore may constitute support for the larc-abc theory.  

	In my case, at least, the core intuition behind 
the desire for a measurement theory involving jumps to energy eigenstates is that if bound 
electrons jumped to energy eigenstates after absorbing incoming particles, they would carry anything 
correlated with them, such as the centres of mass of other bodies, into suitably classical states.  
For bound states, the energy eigenstates seem to form a natural solution to the basis problem, where 
they do not for unbound states.  
A severe barrier to achieving a `jumps to energy eigenstates' theory has been the expectation that it 
would be impossible to treat the 
bound states separately from the other bodies correlated with them, and therefore one would be forced to 
adopt a theory where the entire correlated state would need to jump to an energy eigenstate, even though 
this would include the centre of mass states of free bodies.  

	From this point of view, the crucial element 
of larc theory is that the creation and annihilation operators in larc events provide a barrier that separates 
the bound energy eigenstates from other elements of the total system, so that it is possible for bound 
electrons to jump to energy eigenstates, and carry other systems that are correlated with them into states 
that need not themselves be energy eigenstates.  Nor does the total state of the whole system need to be 
an energy eigenstate.

	The measurement problem is one of the oldest open questions in quantum theory.  The list of attempted 
solutions to this problem is long.  While many have their good points, most have failed, or at least failed to 
be convincing.  Some remain in play, but none that can claim to be a clear solution to the problem.  

	The theory described here has a number of advantages that I believe have not been seen before in a single theory.  
It solves the problem as a matter of straightforward physics, without requiring conceptual contortions in attempting 
to generate the Born rule probabilities, as the Everett theory does.  
It produces localisation of macroscopic 
bodies in the most physically natural way, as a direct result of illumination reflected from one body being 
absorbed by other bodies.  
And it is that rarity, a reduction theory which is naturally relativistic.  

I would like to thank Adrian Flitney, Michael Hall, and Angas Hurst, for valuable discussions and 
criticism.

\end{document}